\documentclass[pra,showpacs,onecolumn,superscriptaddress]{revtex4}
\usepackage{graphicx}
\usepackage{epsfig}
\usepackage{dcolumn}
\usepackage{bm}

\newcommand{\ketbra}[2]{| #1 \rangle \langle #2 |}

\begin{document}

\title{ Variational quantum tomography with incomplete information by means of
semidefinite programs}
\author{Thiago O. Maciel} 
\email{maciel@gmail.com}
\author{Andr\'e T. Ces\'ario}
\email{andretanus@gmail.com}
\author{Reinaldo O. Vianna}
\email{reinaldo@fisica.ufmg.br}
\affiliation{Departamento de F\'{\i}sica - ICEx - Universidade Federal de Minas Gerais,
Av. Ant\^onio Carlos 6627 - Belo Horizonte - MG - Brazil - 31270-901.}

\date{\today}

\begin{abstract}
 We introduce a new method to reconstruct unknown quantum states out
of incomplete and noisy information. 
The method is a linear convex optimization problem, therefore with a unique
minimum, which can be efficiently solved with Semidefinite Programs.
Numerical simulations indicate that the estimated state does not overestimate
purity, and neither the expectation value of  optimal entanglement witnesses.
The convergence properties of the method are similar to {\em compressed 
sensing} approaches, in the sense that, in order to reconstruct
low rank states,  it needs just a fraction of the effort
 correspondig to an informationally complete measurement.

\end{abstract}

\maketitle

\section{Introduction}

All we can know about a quantum  system can be compactly represented by its 
state or density matrix, i.e., a trace one positive operator
($\rho$). The observable properties of the system are represented
by Hermitian operators ($O$), and their expectation values are given
by the rule: $Tr(\rho O)$. 
Consider the system is defined on a Hilbert space of dimension $d$. 
In this case, the state ($\rho$)  is a $d\times d$ Hermitian matrix,
 or a real $d^2$ dimensional vector in the Hilbert-Schmidt space. 
As the state is normalized
($Tr(\rho)=1$), it has just $d^2-1$ independent  real parameters, so it is 
in fact a $d^2-1$ dimensional real vector. Therefore, a carefully chosen
 set of $d^2-1$ independent measurements, in the sense  that it forms a complete
basis in the Hilbert-Schmidt space,  allows for the reconstruction of the state.
This approach is usually referred to as quantum tomography.

There are many  methods to estimate a quantum state (for a review, see, 
for example, \cite{QSE}).
 An important class of these methods is known
as {\em maximum-likelihood quantum tomography} (MaxLik) (see, for example,
\cite{ML0,ML1,ML2,ML3}). MaxLik 
is a statistical inference approach which yields  probability distributions
 as close as possible to
the measured frequencies.
Nonetheless, these probability distributions frequently correspond to non-positive
operators, due to the always present  experimental imprecisions.   It is important to stress that our method
does not follow this line. Our heuristic consists of performing a  variational search 
in the state space for the best trace one positive operator compatible with the measured
data. 
Note also that MaxLik cannot guarantee that fake entanglement 
be not attributed to the system \cite{Horodecki}.
Our numerical simulations indicate that the method we propose does not
overestimate entanglement.
Another important aspect of our method is that it can be solved
exactly, for it can be cast as a linear convex optimization
problem, while the usual algorithms for quantum state
tomography, though convex,  are non-linear,  and the actual
algorithms  implementing them can be plagued with many
local minima.

To perform the complete set of measurements
necessary to reconstruct the quantum state 
 of  a moderately sized system 
 could be impossible in practice.
As an example, consider 
the experimental generation of the 8-qubit entangled state 
 reported by H\"affner {\em et al.} \cite{Blatt}. 
 To tomograph the  8-qubit system,  65536 projectors needed to be measured ($2^{16}$), 
and the computational effort to apply MaxLik to such a 
state is discouraging, though it was done. But the tomography of a larger system
could be inaccessible experimentally, and yet we would like to assess its properties.
The exponential growth of the Hilbert space is insurmountable, 
in this respect. 
Using the  Maximum Entropy 
Inference Principle of Jaynes \cite{Jaynes} (MaxEnt), one can estimate a quantum 
state out of incomplete information.
In this approach, one searches for the maximum entropy state,
compatible with the measured data \cite{Buzek}.
This state corresponds to maximum ignorance with respect
to the unmeasured set of observables forming the complete basis for the density 
operator.
The Horodecki \cite{Horodecki}
showed an example where the direct application of MaxEnt yielded fake
entanglement, i.e., the reconstructed state was entangled, while the
available data were compatible with a non-entangled state.
Our method can also estimate a state out of incomplete information, 
and  with the advantage of not overstimating entanglement, as indicated
by our simulations.

Another interesting aspect of  our method is that it can  identify {\em incompatible data}.
Suppose  a set of complete measurements are being performed, aiming  a 
quantum state reconstruction. It can happen that, for a particular complete
measurement,
some parameters of the experimental setup have fluctuated to a point of
changing the quantum state, but it went unnoticed by the experimentalist.  
In this case, data for different states have been collected. 
When our method is applied to these measurements, the algorithm will
detect that the data are incompatible with a single quantum state.
Then some of the  data can be discarded, and the state  can be reconstructed.

In the next section,  we  develop a variational method to
estimate a state, which  relies on  incomplete and noisy information
of the quantum state. 
The method we obtain has the particular form of  a linear convex optimization problem, 
known as Semidefinite Program (SDP), for which 
efficient and stable algorithms are
available \cite{Boyd, sedumi, yalmip}. Though this unfortunate name
can suggest a kind of black-box computational algorithm, a SDP is not
so. It consists of minimizing a linear objective under a linear matrix
inequality constraint, precisely,
\begin{center}
{\em minimize} $c^\dagger x$
\begin{equation}
 subject\,\, to \left\{ F(x)=F_0+\sum_{i=1}^m x_iF_i\geq 0, \right.
\end{equation}
\end{center}
where $c\in C^m$ and  the Hermitian matrices $F_i\in C^{n\times n}$ are
given, and $x \in C^m$ is the vector of optimization variables.
$F(x)\geq 0$ means that $F(x)$ is a positive matrix. The problem defined
in Eq.1 has no local minima. When the unique minimum of this problem 
cannot be found analytically, one can resort to powerful algorithms that
return the exact answer \cite{sedumi}. To solve the problem in Eq.1 could be
compared to finding the eigenvalues
of a Hermitian matrix. If the matrix is small enough or has very high
symmetry, one can easily determine its eigenvalues, but in other cases
some numerical algorithm is needed. Anyway, one never doubts that the
eigenvalues of such a matrix can be determined exactly.
We point out that we  have successfully used SDPs before, in the development of  
powerful methods to construct entanglement witnesses \cite{rov1,rov2,rov4,rov3}.

After discussing the methodology, we present a section with numerical examples that we consider
representative, namely, a full rank highly symmetric two-qutrit density matrix, and a
pure (rank one) five-qubit state.  
Our method works equally well in 
these two extreme cases.  Then we show reconstruction of four-qubit
mixed states of all ranks, in order to have a better taste
of the convergence properties of our method. These numerical examples
suggest that our method converges very fast for low rank states, or
for states that, though of high rank, have very high symmetry. 
These convergence properties are similar to  recent introduced
approaches in {\em compressed sensing} \cite{Eisert}. 
To reconstruct, with high fidelity,  full rank states with no symmetry,
an informationally complete measurement is needed. But even in these  
difficult cases,
our method can yield reasonable lower bounds to entanglement out of incomplete information \cite{rov3}.
We illustrate this feature by evaluating  optimal entanglement witnesses 
\cite{rov1,rov2,rov4,rov3}in qubit-qutrit full
rank states, with increasing number of measurements.
Besides the numerical examples, we have also successfully tested our approach
in a real experiment, where entangled qutrits were generated in a quantum
optics setup \cite{Lima}.

\section{Theory}

The  state  of a quantum system is represented by its density matrix $\rho$,
which is  a $d^2-1$ dimensional real vector in the Hilbert-Schmidt space.
 Therefore, we can write it as:
\begin{equation}
\label{rho3}
\rho=C_0 I +\sum_{l=1}^{d+1}\sum_{i=1}^{d-1} C_{il}P_{i}^{l} =C_0 I + \sum_{\lambda=1}^{d^2-1} C_\lambda P_\lambda ,  
\end{equation}
where   the $P_i^l=P_\lambda$  are Hermitian operators, forming
a complete basis in the Hilbert-Schmidt space, $I$ is the identity,
 and $\lambda$ is just
a convenient compact index.

Now, for concreteness,  we assume the 
operators 
in each {\em class} $l$ are rank one projectors, forming a
complete measurement, i.e.,
\begin{equation}
\label{mub1}
\begin{array}{l}
I=\displaystyle{\sum_{i=1}^{d}}P_i^l , \\
 \\
Tr(P_i^lP_j^l)=\delta_{ij}.
\end{array}
\end{equation}
$C_0$, which is not an independent parameter, in this case is given by
$C_0=(1-\sum_{\lambda=1}^{d^2-1} C_\lambda)/d$. Note that, in Eq.\ref{rho3}, we use
just $d-1$ projectors from each complete measurement labelled by $l$
(for the $d$ probabilities sum to one). Besides being convenient
for our calculations, this basis also  has the minimum number of projectors
necessary to expand a Hermitian matrix with fixed trace, but the method
we derive is not basis dependent. For instance, if one chooses  to expand
the state in a SU basis, like a generalized Bloch representation, 
\begin{equation}
\rho=\frac{1}{d}I+ \sum_{i=1}^{d^2-1} r_i \sigma_i\, ,
\end{equation}
to  obtain the coefficients $r_i=Tr(\rho \sigma_i)$, 
one should consider the spectral decomposition 
 $\sigma_i=\sum_{j=1}^d \alpha_{ij} \ketbra{i_j}{i_j}$,
and we are back to the measurement of projectors, 
$Tr(\ketbra{i_j}{i_j}\rho)$. But now, instead of the $d^2-1$ projectors
of Eq.2, one has  $d\times (d^2-1)$ projectors, i.e., one
complete measurement for each operator $\sigma_i$. The use of POVMs 
(Positive  Operator Valued Measure) also poses no difficulties, for
it consists of measurements of  rank one projectors again.

We assume that from a set of $d^2-1$ projectors (or observables), forming
a complete basis in the Hilbert-Schmidt space (the space of linear operators),
just $N (< d^2-1)$ have been measured. We refer to measured and unmeasured 
projectors (observables) as the {\em known} and {\em unknown} sets, respectively.
Now we introduce the following  {\em cost operator} (the sum of the projectors in the unknown set), which can be thought of as a {\em Hamiltonian}:
\begin{equation}
H=\sum_{\lambda=N+1}^{d^2-1} P_\lambda.
\end{equation}
We want a trace one positive operator, which minimizes the {\em cost function}:
\begin{equation}
E=Tr(H\tilde{\varrho})=\sum_{\lambda=N+1}^{d^2-1} q_\lambda,
\end{equation}
where $q_\lambda$ are positive numbers corresponding to the projections
of $\tilde{\varrho}$ on the unknown set.
Our variational principle now reads:
\begin{equation}
\delta E=Tr(H\delta\tilde{\varrho})=0.
\end{equation}
The Lagrangian for this problem reads:
\begin{equation}
\begin{array}{ll}
\label{L1}
L= & E+\alpha[Tr(\tilde{\varrho})-1]+\displaystyle{\sum_{\lambda=1}^{N}}\beta_\lambda[Tr(\tilde{\varrho}P_\lambda)-p_\lambda]   +  \\ &
\displaystyle{\sum_{\lambda=N+1}^{d^2-1}}\gamma_\lambda[Tr(\tilde{\varrho}P_\lambda)-q_\lambda],
\end{array}
\end{equation}
where $\alpha$, $\beta_\lambda$, and $\gamma_\lambda$ are Lagrange 
multipliers corresponding to the constraints on the quantum state.

 Eq.2 in matrix form reads
\begin{equation}
\rho=C^T P,
\end{equation}
where $C$ is a column vector with the real coefficients $C_\lambda$, and
$P$ a column vector collecting the matrices $P_\lambda$.
Consider the overlap matrix $S$, with elements $S_{\mu \nu}=Tr(P_\mu P_\nu)$, and the column vector $\mathcal{P}$, with elements $p_\lambda=Tr(\rho P_\lambda)$. Then we have
\begin{equation}
C=S^{-1}\mathcal{P}.
\end{equation}
Only when the probabilities in $\mathcal{P}$ are exact, the vector $C$ yields a positive operator, by means of Eq.9. But it is never the case, in
practice. The frequencies obtained from an experiment  
are noisy, thus:
\begin{equation}
\label{pexp}
Tr(\rho P_\lambda) \in [p_\lambda-\varepsilon,p_\lambda+\varepsilon],
\end{equation}
with $\varepsilon$ positive, and hopefully small.

To account for Eq.\ref{pexp} in our algorithm, we introduce the additional 
constraints:
\begin{equation}
\label{erro1}
\begin{array}{l}
Tr(\tilde{\varrho}P_{\lambda}) \geq (1-\Delta_{\lambda})p_\lambda,\\
Tr(\tilde{\varrho}P_{\lambda}) \leq (1+\Delta_{\lambda})p_\lambda,\\
\Delta_\lambda\geq 0, \,\,\, \lambda \in [1,N].
\end{array}
\end{equation}
Again, we want to determine a trace one positive operator $\tilde{\varrho}$, with
minimum $\Delta_\lambda$.
It is important to note that Eq.12 is not to be interpreted as any kind of statistical treatment.
By minimizing $\Delta_\lambda$, we are simply adjusting the frequencies obtained in the laboratory, such that we get a positive matrix by
means of Eqs.9 and 10.  
This works only when Eq.11 is valid for all frequencies. If some
frequencies came from another state, say $Tr(\rho_{other}P_\lambda)$,
then the algorithm will diverge, and we have identified {\em incompatible data}.
To build an appropriate Lagrangian for this
problem, we introduce positive numbers $v_\lambda$ and $w_\lambda$, which
should be minimized, 
and rewrite Eq.\ref{erro1} as:
\begin{equation}
\label{erro2}
\begin{array}{l}
Tr(\tilde{\varrho}P_{\lambda}) - (1-\Delta_{\lambda})p_\lambda = v_\lambda,\\
(1+\Delta_{\lambda})p_\lambda -  Tr(\tilde{\varrho}P_{\lambda})= w_\lambda.
\end{array}
\end{equation}
The Lagrangian now reads:
\begin{equation}
\label{L2}
\begin{array}{ll}
L= & E+
\displaystyle{ \sum_{\lambda=1}^N}(\Delta_\lambda + v_\lambda + w_\lambda) +
  \alpha[Tr(\tilde{\varrho})-1]+ \\ &
\displaystyle{\sum_{\lambda=N+1}^{d^2-1}}\gamma_\lambda[Tr(\tilde{\varrho}P_\lambda)-q_\lambda]+ \\
 & \displaystyle{\sum_{\lambda=1}^{N}}\{\zeta_\lambda[(1+\Delta_{\lambda})p_\lambda -  Tr(\tilde{\varrho}P_{\lambda})- w_\lambda] +  \\ &
\eta_\lambda[  Tr(\tilde{\varrho}P_{\lambda}) - (1-\Delta_{\lambda})p_\lambda - v_\lambda]\},
\end{array}
\end{equation}
where $\zeta_\lambda$ and $\eta_\lambda$ are additional Lagrange multipliers.
Note that this Lagrangian defines a convex problem. We want to minimize
a linear function, constrained by matrix inequalities. Our variable is
$\tilde{\varrho}$, and the search space is restricted to the
cone of positive matrices. This is a typical convex optimization problem,
known as Semidefinite Program (SDP), as described in the Introduction (Eq.1). SDPs have  a unique minimum, and can
be solved very efficiently \cite{Boyd}. Our SDP reads:
\begin{center}
{\em minimize}  $(E+\sum_{\lambda=1}^{N} \Delta_\lambda)$ 
\begin{equation}
\label{sdp2}
 subject\,\, to
\left\{
\begin{array}{l}
 \tilde{\varrho} \geq 0, \\
 Tr(\tilde{\varrho})=1,  \\  
\Delta_\lambda \geq 0, \\
 Tr(\tilde{\varrho} P_{\lambda}) \geq (1-\Delta_\lambda) p_{\lambda}, \\
 Tr(\tilde{\varrho} P_{\lambda}) \leq (1+\Delta_\lambda) p_{\lambda}, \,\,\,
 \forall \lambda \in [1,N]\, . 
\end{array}
\right.
\end{equation}
\end{center}
Now we have concluded the derivation of our method. Eq.\ref{sdp2} returns a 
state $\tilde{\varrho}$ which is the optimal approximation to the
unknown state $\rho$.

\section{Applications}

\begin{figure}
\includegraphics[scale=0.41]{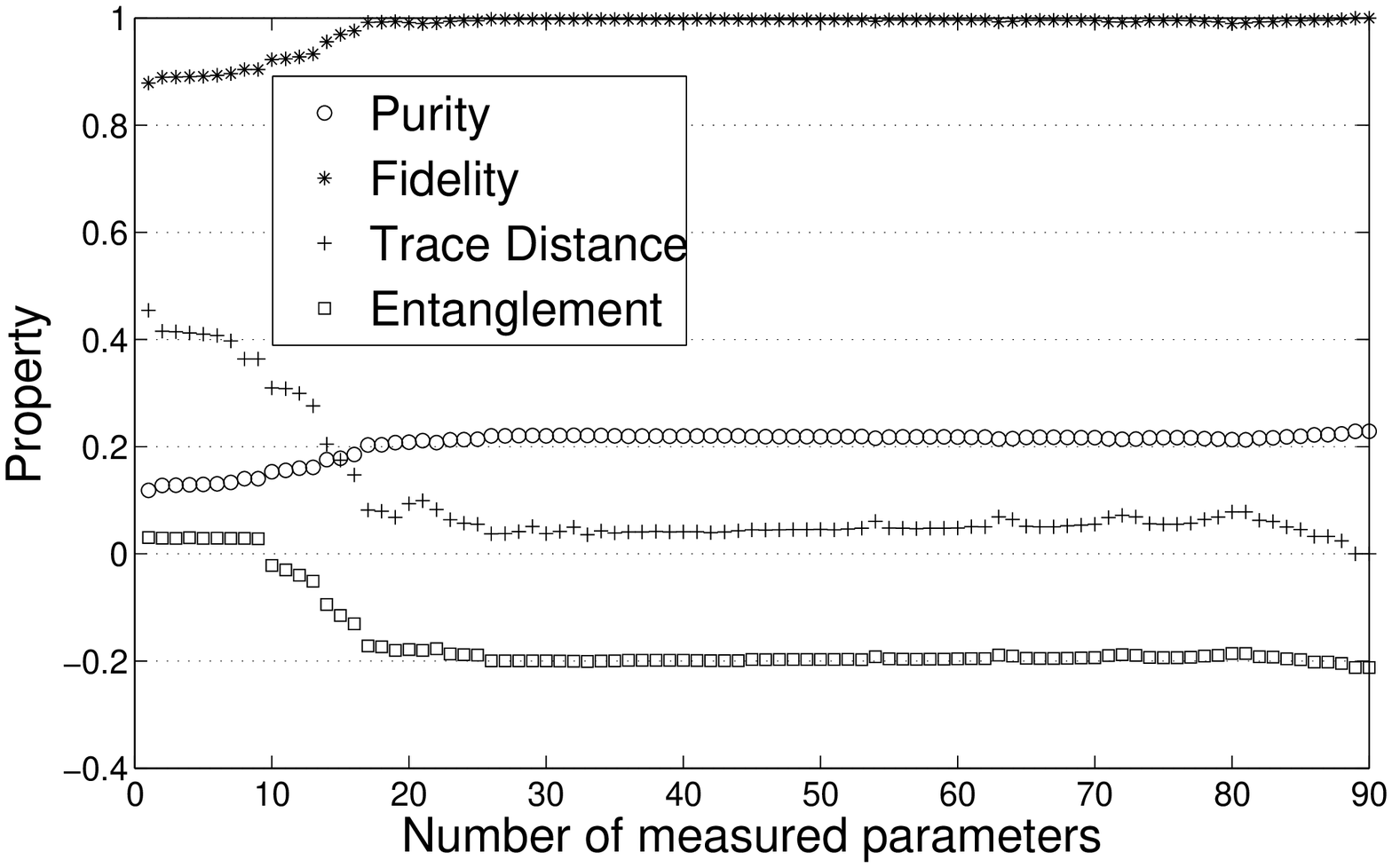}
\includegraphics[scale=0.41]{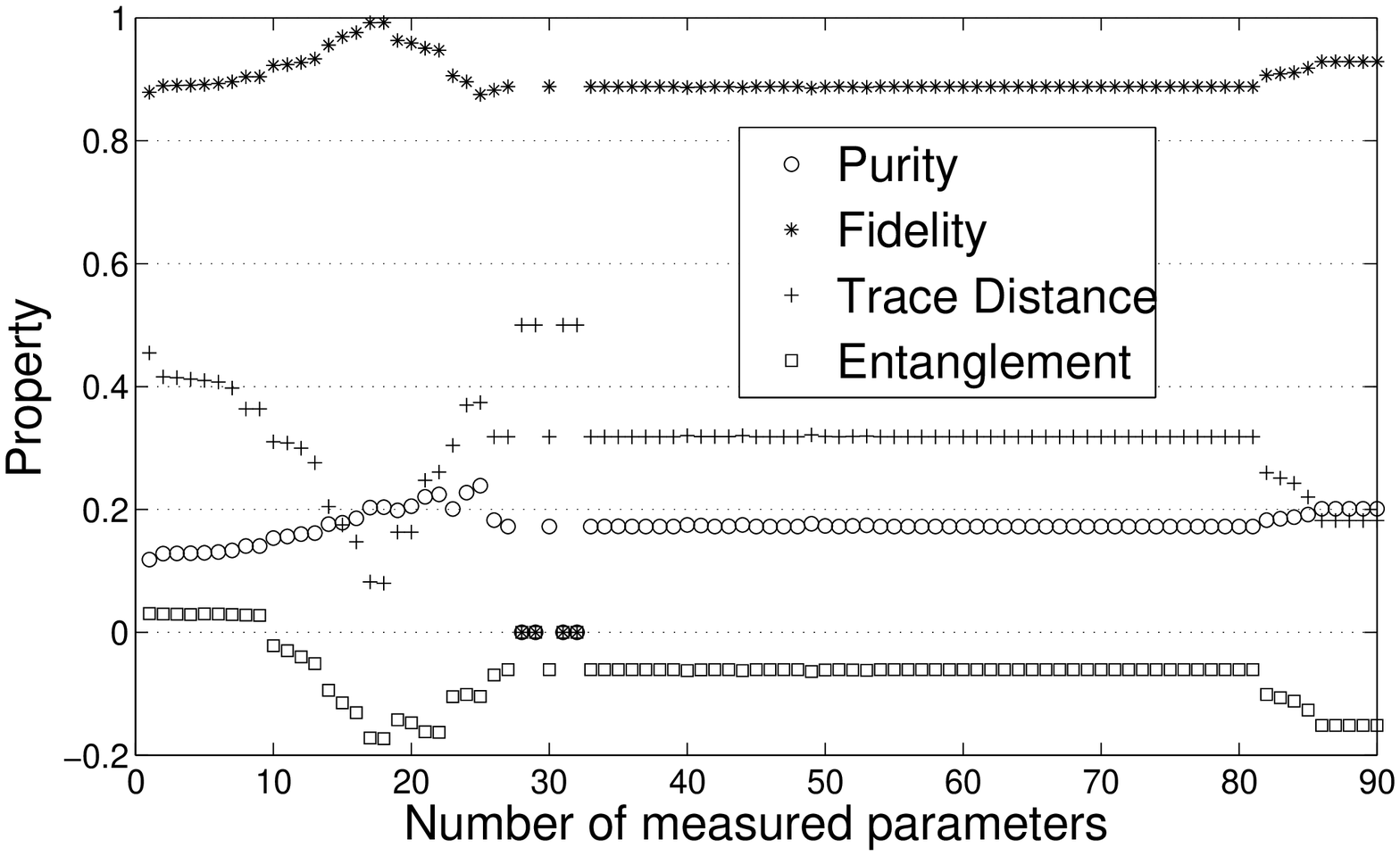}
\includegraphics[scale=0.41]{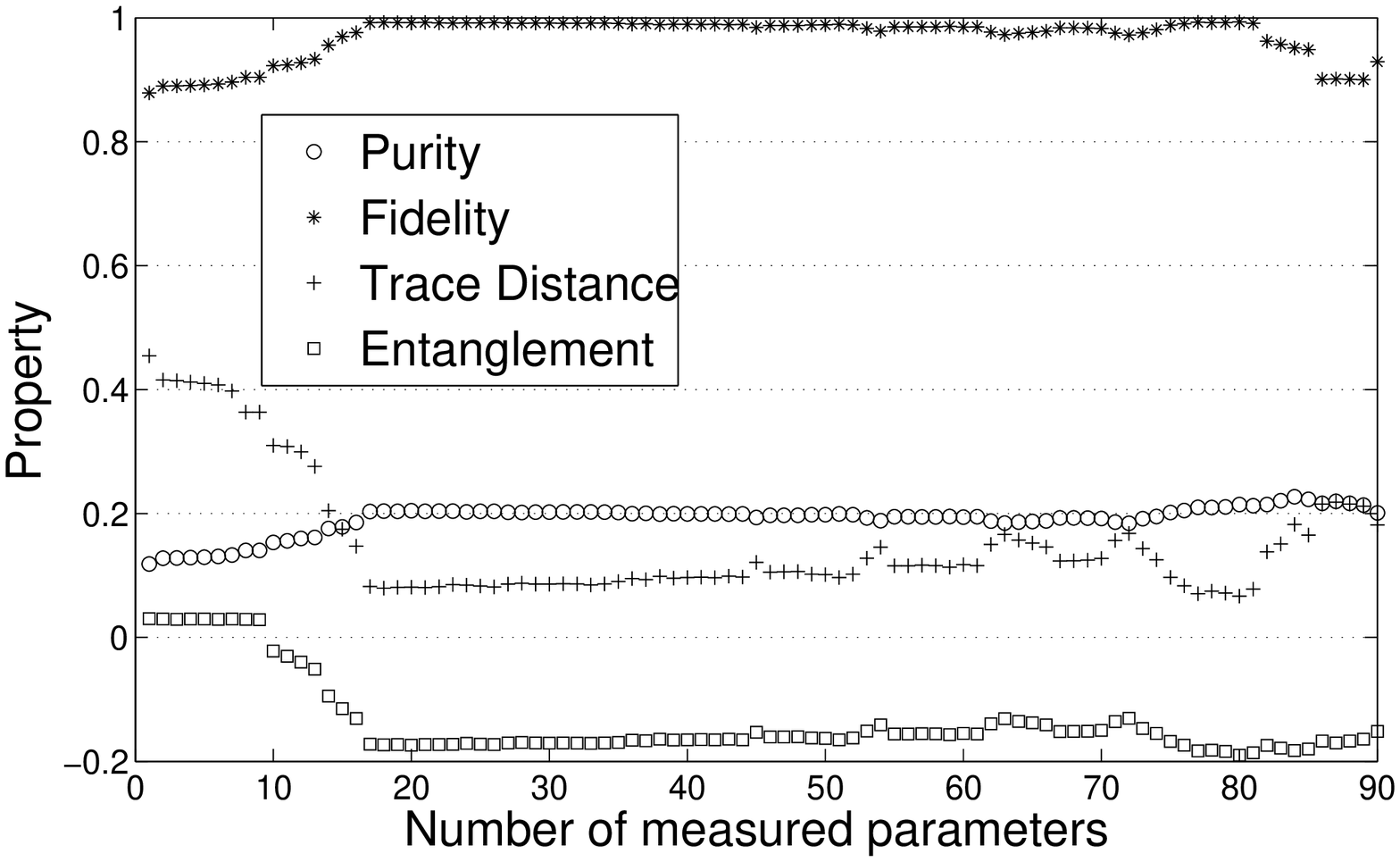}
\caption{Estimation of some properties and reconstruction of a 
two-qutrit full rank Werner state ($\beta=-0.8$, {\em cf.} Eq.16).
 We plot Purity, $Tr(\tilde{\varrho}^2)$, Fidelity,
 $Tr(\sqrt{\rho^{\frac{1}{2}}\tilde{\varrho}\rho^{\frac{1}{2}}})$,
Trace Distance, $\frac{1}{2}Tr|\tilde{\varrho}-\rho|$, and entanglement,
$Tr(W_{\tilde{\varrho}} \tilde{\varrho}  )$ ($W_{\tilde{\varrho}}$ is the trace-one optimal entanglement witness),
 against the number of measured projectors.
The {\em measurements} have statistical errors of up to 50\%, according
to a uniform distribution. In the last two panels, we introduced
measurements from another Werner state ($\beta=+0.8$), to show that
the algorithm is capable to identify {\em incompatible data} (see text
for details).      }
\end{figure}

\begin{figure}
\includegraphics[scale=0.41]{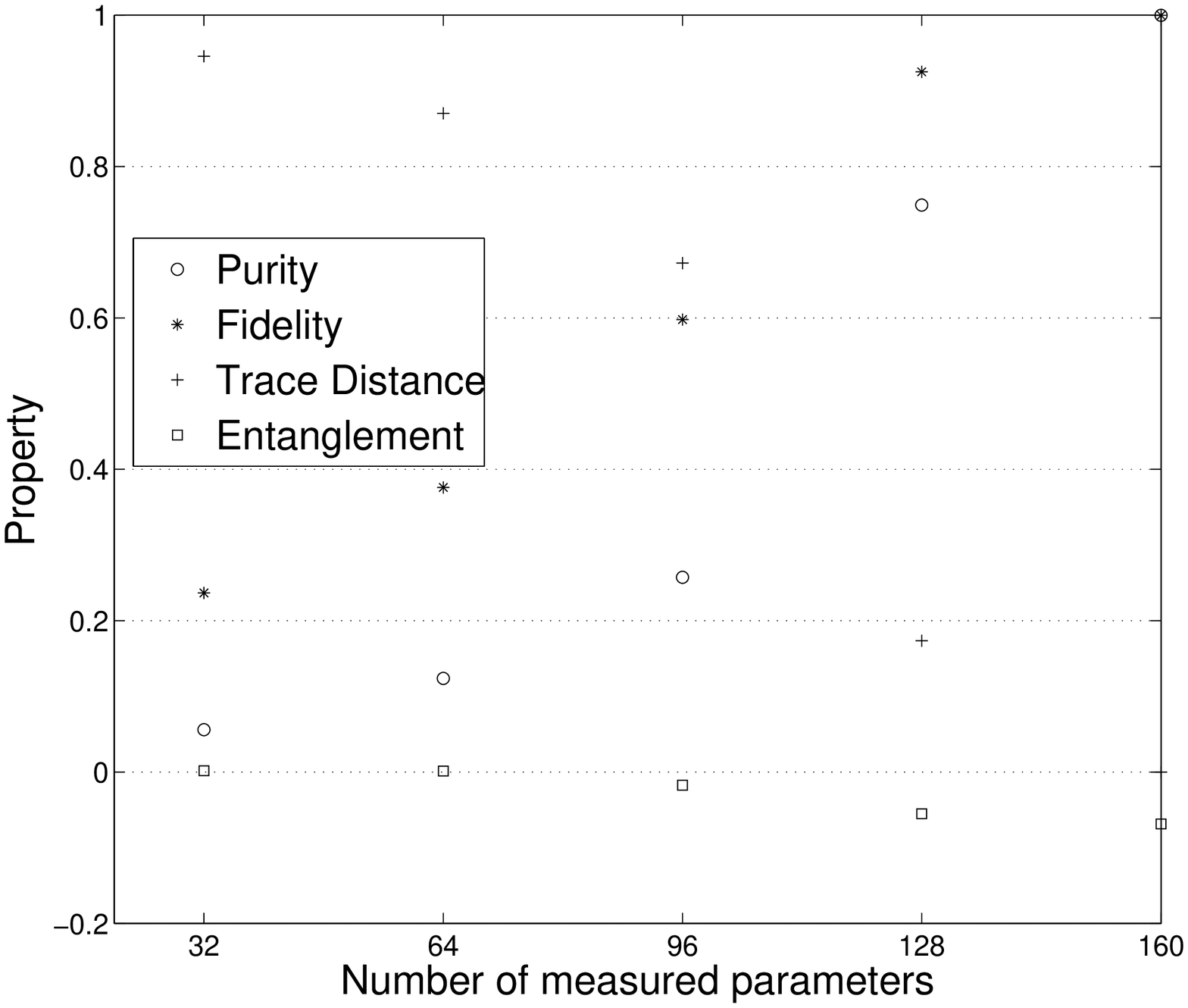}
\caption{Reconstruction of a five-qubit low entangled pure state.
 We plot Purity, $Tr(\tilde{\varrho}^2)$, Fidelity,
 $Tr(\sqrt{\rho^{\frac{1}{2}}\tilde{\varrho}\rho^{\frac{1}{2}}})$,
Trace Distance, $\frac{1}{2}Tr|\tilde{\varrho}-\rho|$, and entanglement,
$Tr(W_{\tilde{\varrho}} \tilde{\varrho}  )$ ($W_{\tilde{\varrho}}   $ is the trace-one optimal entanglement witness),
 against the number of measured projectors.
 Note that
the exact state is recovered with just 160 measurements out of   the total of
1056, that would be necessary to completely span the Hilbert-Schmidt space.}
\end{figure}

Now we would  like to 
illustrate the use of Eq.\ref{sdp2}. As the examples 
show,  our method does not overestimate purity and neither the
expectation value of optimal entanglement witnesses \cite{rov1, rov2, rov4, rov3},
and it can also  identify incompatible data.
The examples also show 
that few measurements are needed to reconstruct both low rank states,
 and full rank states with high symmetry (Figs. 1, 2 and 3). 
In Figs. 1, 2 and 3,  the set of measurements  we chose for the state expansion is  mutually unbiased bases (MUB) \cite{Ivanovic, Wootters}, 
in the sense that any
two vectors of different complete measurements  have the same overlap's absolute value. As demonstrated by Wootters {\em et. al} \cite{Wootters} and
Ivanovic \cite{Ivanovic}, this is the best informationally complete 
projective measurement one can do. It is optimal both in the statistical sense
and in the number of projectors to be measured. The POVMs that would be
equivalent to MUBs are the Symmetrically Informationally Complete POVMs 
(SIC-POVM) \cite{povm}. But while the MUBs are known for all Hilbert spaces
which have dimension of  power of a prime, SIC-POVMs, though conjectured
to exist in all dimensions, are known just in a few particular cases. 
In Fig. 4, we used the SU(6) basis, to illustrate that our method works
with any kind of basis.

Our first example is 
a highly mixed 2-qutrit Werner state \cite{Werner}
 (purity=0.23), which is full rank, and
has entanglement -0.21, according to its optimal trace one entanglement
witness \cite{rov1,rov2}. The explicit expression of this state is:
\begin{equation}
\label{wern}
\rho=\frac{I+\beta  F}{9+3 \beta},
\end{equation}
with $-1\leq \beta \leq 1$. $\rho$ is separable for
$\beta\geq-\frac{1}{3}$.
$F$ is a swap operator for two qutrits,
\begin{equation}
 F= \sum_{i,j=1}^{3}\ketbra{ij}{ji} .
\end{equation}
    For two qutrits, we have ten complete measurements, with nine
projectors in each one.
 In Fig.1, we plot,
 against the number of measured projectors, 
 purity,
 fidelity and trace distance
to the true or exact state,
and the witnessed entanglement ($Tr(W_{\tilde{\varrho}}\tilde{\varrho} )$ - the more
negative is this expectation value, the more entangled is the state).
In the calculations of entanglement,
we use an optimal trace one witness, obtained with techniques developed
in \cite{rov1,rov2,rov4,rov3}.  The optimal entanglement witness  has to be calculated
for each state individually.
The  (simulated) measured frequencies (Eq.11) have
statistical errors of up to 50\%, according to a random uniform distribution.
In the first panel, one can see that with about twenty measurements,
from a total of ninety, we have a practically perfect estimation of
the entanglement, and the state was reconstructed with high fidelity, 
or low trace distance. Note that all the properties we are calculating 
converge monotonically, never been overestimated. In the second panel,
the third complete measurement (indexed 19 to 27) corresponds to a 
different Werner state. Note that we had convergence up to the 
$18^{th}$ projection, and then all the properties started to diverge,
to resume convergence again in the fourth measurement. In the last panel,
the {\em incompatible data} was moved to positions 81 to 90, and one can see convergence
for the measurements before this. The fluctuations seen in all the panels
are due to the large statistical errors we imposed, added to the fact that without
all the complete measurements, there is always the possibility of a family of
close states to fit the data.

To highlight the efficiency of our method, in Fig.2 we applied it to a
a five-qubit low entangled pure state (it is a normalized random vector with
32 complex coefficients). In principle, one needs to perform
33 complete measurements (1056 projectors) to reconstruct such a state,
but we needed just five (160 projectors). The entanglement plotted in the
figure is the genuine five-party one, given by the trace one optimal
entanglement witness \cite{rov1,rov2,rov4,rov3}. The tomography calculation  took six 
seconds in a laptop, with a 2.8GHz processor, and 3GB of RAM, 
running MATLAB under Linux. The five-party entanglement calculation 
was more expensive, and took about forty minutes.

Fig.3 illustrates the convergence properties of our method, as the
rank of the density matrices increases. For a system of four qubits 
(Hilbert-Schmidt space dimension of 256), we consider one hundred
random density matrices of each rank, varying from pure states (rank one) to
full rank states (rank 16). We plot the average number of measurements to
reconstruct the state with high fidelity ($Tr|\rho-\tilde{\varrho}|<10^{-6}$), against the rank. In the 
figure we also indicate the minimum (Best Case) and maximum (Worst Case)
number of measurements needed in the reconstruction, in each sample.
We see that the state reconstruction needs very few measurements for low 
rank states, and it can need all the measurements for high rank states.
This behavior is similar to the {\em compressed sensing} approach
reported in \cite{Eisert}.

\begin{figure}
\includegraphics[scale=0.41]{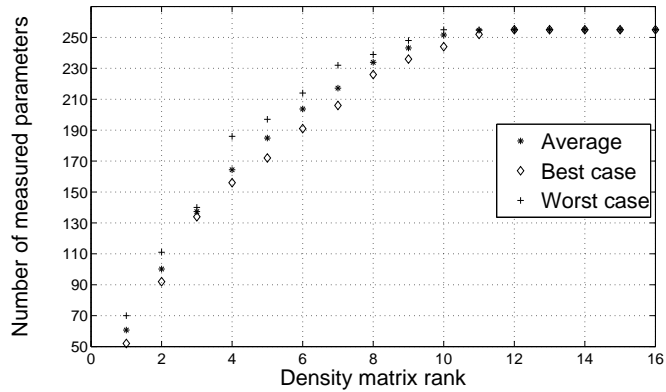}
\caption{Reconstruction of  random four-qubit entangled states of all ranks.
The average number of measurements needed to a highly  faithful
reconstruction ($Tr|\rho-\tilde{\varrho}|<10^{-6}$) is plotted against the rank of the state.
For each rank, it is employed a sample of one hundred states. For each
sample, it is also plotted the minimum (Best Case) and maximum (Worst Case)
number of measurements needed.}
\end{figure}

As a last example, we consider a qubit-qutrit system. The states will be
expanded in the 36  SU(6) observables ($\sigma_i$) 
 forming  an informationally  complete
basis in the Hilbert-Schmidt space ({\em cf.} Eq.4). 
The {\em Hamiltonian} ({\em cf.} Eq.5) to be used in Eq.15 is formed
by the eigenprojectors of the $36-N$ unmeasured $\sigma_i$ ({\em cf.} Eq.4).
For $10^3$ random full-rank density matrices ($\rho_i$), we plot, against the number of
measured observables, the average trace distance
($10^{-3}\sum_{i=1}^{1000} Tr|\tilde{\varrho_i}-\rho_i|/2$), and average fraction
of entanglement, defined as follows. The entanglement of a state 
is $E(\rho)=Tr(W_\rho \rho)$, if this expectation value is negative,
and zero otherwise. $W_\rho$ is the optimal trace one entanglement witness
for the state $\rho$, calculated with the techniques described in \cite{rov1,
rov2,rov4,rov3}.  Thus, the {\em Entanglement} plotted in the figure is given
by $10^{-3}\sum_{i=1}^{1000} E(\tilde{\varrho_i})/E(\rho_i)$.
The full rank states will need all the 36 measurements
to be reconstructed with high fidelity, for they have no symmetry. Even though, Fig.4 shows that
 a non null lower bound
to the entanglement is obtained with less measurements, and of course this bound
tends to the correct entanglement as the number of measurements increases.

\begin{figure}
\includegraphics[scale=0.41]{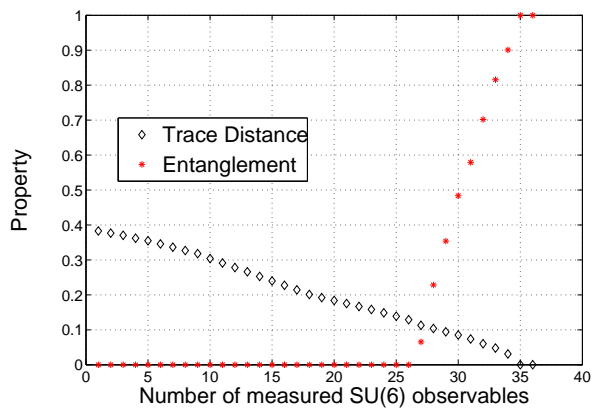}
\caption{Reconstruction of $10^3$ full rank   random  qubit-qutrit  entangled states.
We plot the fraction of the true entanglement, on average, as the number
of measurements increases (see text for details). The average trace
distance to the {\em true} state is also plotted. 
We see that an almost  perfect reconstruction is possible only with all 
the measurements, but non null lower bounds to the entanglement are
obtained with less measurements. }
\end{figure}

\section{Conclusion}

We developed a method that yields an estimate ($\tilde{\varrho}$) of a 
unknown quantum state ($\rho$),
without the need to perform an informationally complete measurement.
Our numerical simulations indicate that our method does not overestimate
entanglemnt, and does not underestimate purity. These quantities
tend to the true values, as the number of measurements
increases towards an informationally complete measurement.
Low rank states, or high rank states with high symmetry can 
be reconstructed with high  fidelity, using few measurements.
This is simply because the number of independent parameters in
such density matrices is much less than $d^2$, the dimension of
the Hilbert-Schmidt space. 
 Note that it is a
line of investigation we started in \cite{rov3}, in the context of
entanglement detection with few measurements. 
The method can be useful in the study of larger systems, where an
informationally complete measurement is out of question.
In this respect, we note that the convergence properties of our method
is similar to the {\em compressed sensing} approach recently introduced
in \cite{Eisert}.

We conclude by mentioning that our method has been successfully 
employed in a real experiment reported in \cite{Lima}.
We also mention that we have extended our method to the problem
of process tomography \cite{processo}.

{\em Acknowledgments} - We thank C.H. Monken, S.P. Walborn and 
P.H Souto Ribeiro for the discussions. 
We also acknowledge Fernando G.S.L. Brand\~ao and the referee for pointing
out a mistake, in the manuscript,  concerning lower bounds of expectation values.

Financial support by the
Brazilian agencies  FAPEMIG, and  INCT-IQ (National 
Institute of Science and Technology for Quantum Information).


\begin{thebibliography}{99}

\bibitem{QSE}
M. Paris, J. Rehacek (Eds), {\em Quantum State Estimation}., Lect. Notes
Phys. {\bf 649} (Springer, Berlin Heidelberg 2004), DOI 10.1007/b98673.

\bibitem{ML0} 
Z. Hradil, J. Rehacek, J. Fiurasek, M. Jezek, ``Maximum-Likelihood 
Methods in Quantum Mechanics'', in M. Paris, J. Rehacek (Eds), {\em Quantum State Estimation}., Lect. Notes
Phys. {\bf 649} (Springer, Berlin Heidelberg 2004), DOI 10.1007/b98673,
pp. 59-100.

\bibitem{ML1}
G.M. D'Ariano, D.F.   Magnani,  P.  Perinotti,  Phys. Lett. A,  {\bf 373},
111-115 (2009).

\bibitem{ML2} G.M.  D'Ariano, P.  Perinotti, P., Phys. Rev. Lett., {\bf 98},
020403 (2007).

\bibitem{ML3} J. Rehacek, Z.  Hradil, E.  Knill, A.I.  Lvovsky, 
Phys. Rev. A, {\bf 75}, 042108 (2007).


\bibitem{Horodecki} R. Horodecki, M. Horodecki, P. Horodecki,
Phys. Rev. A, {\bf 59}, 1799-1803 (1999).

\bibitem{Jaynes} E.T. Jaynes, Phys. Rev. {\bf 106}, 620-630 (1957).

\bibitem{Buzek} V. Buzek, ``Quantum Tomography from Incomplete
Data via MaxEnt Principle'', in 
M. Paris, J. Rehacek (Eds), {\em Quantum State Estimation}., Lect. Notes
Phys. {\bf 649} (Springer, Berlin Heidelberg 2004), DOI 10.1007/b98673,
pp. 189-230.





\bibitem{Blatt} H. H\"affner, W.  H\"ansel, C.F.  Roos, J.  Benhelm,
D. Chek-al-kar, M.  Chwalla, T.  K\"orber, U.D.  Rapol, 
M. Riebe,  P.O. Schmidt, C. Becher, O.  G\"uhne, W.  D\"ur,
R. Blatt,  Nature, {\bf 438}, 643-646 (2005).

\bibitem{Boyd}  S.  Boyd, L.   Vandenberghe,   {\em Convex Optimization},
(Cambridge University Press, Cambridge, 2000).

\bibitem{sedumi} J.F.  Sturm,   Optim. Methods and Software {\bf 11},
625-653 (1999); \newline SEDUMI: 
\begin{tt}http://fewcal.kub.nl/sturm/softwares/sedumi.html\end{tt}.

\bibitem{yalmip} 
J. L{\"o}fberg,
 {YALMIP} : A Toolbox for Modeling and Optimization in {MATLAB},
 Proceedings of the {CACSD} Conference,
 2004,
Taipei, Taiwan, (unpublished) \begin{tt}http://control.ee.ethz.ch/\~{}joloef/yalmip.php\end{tt}.  


\bibitem{rov1} F.G.S.L Brand\~ao,  and R.O.  Vianna,  Phys. Rev. Lett.
{\bf 93}, 220503 (2004).


\bibitem{rov2} F.G.S.L.  Brand\~ao,  and R.O.  Vianna, R.O.,  Int. J. Quantum Inf.
{\bf 4}, 331-340 (2006).

\bibitem{rov4}  F.G.S.L.  Brand\~ao,  and R.O.  Vianna,  Phys. Rev. A, {\bf 70},
062309 (2004).

\bibitem{rov3} T.O. Maciel,  and R.O. Vianna,  Phys. Rev. A, {\bf 80}, 
032325 (2009).



\bibitem{Eisert} D. Gross, Y.  Liu, T.  Flammia, T., S.  Becker, J. Eisert,
Phys. Rev. Lett. {\bf 105}, 150401 (2010).


\bibitem{Lima}  G. Lima, E.S G\'omez, A. Vargas, R. O. Vianna, and
C. Saavedra, Phys. Rev. A, {\bf 82}, 012302 (2010).


\bibitem{Werner} R.F.  Werner,  Phys. Rev. A {\bf 40}, 4277-4281 (1989).


\bibitem{Ivanovic} I.D.   Ivanovic,   J. Phys. A: Math. Gen. {\bf 14}, 3241
(1981).

\bibitem{Wootters}  W.K. Wootters,  B.D.  Fields,  Ann. Phys. {\bf 191},
363-381(1989).

\bibitem{povm} J. K. Renes, R. Blume-Kohout, A. J. Soctt, C. M. Caves,
J. Math. Phys. 45, 2171 (2004).


\bibitem{processo} 
T.O. Maciel and  R.O. Vianna,
{\em    Optimal estimation of quantum processes using incomplete information: variational quantum process tomography}, 
arXiv:1007.2395.




\end{thebibliography}
\end{document}